\documentclass{article}

\usepackage{yfonts}
\usepackage{amsmath,amssymb,amsthm}
\usepackage{units,stmaryrd}
\usepackage{graphicx}
\usepackage[all]{xy}
\usepackage{enumerate}
\usepackage[usenames,dvipsnames]{color}
\usepackage[colorlinks]{hyperref}
\hypersetup{
  colorlinks,
  citecolor=Violet,
  linkcolor=Red,
  urlcolor=Blue}
\usepackage{mathtools}

\newcommand{\la}{\langle}
\newcommand{\ra}{\rangle}

\begin{document}

\title{The Selfish Algorithm} 
\author{Eduardo Hermo Reyes \hspace{1.5cm} Joost J. Joosten \\ \small{Department of Logic, History and Philosophy of Science} \\ \small{University of Barcelona}\\ \small{Barcelona, Spain} \\ \href{mailto:ehermo.reyes@gmail.com}{ehermo.reyes@gmail.com} \hspace{1.5cm} \href{mailto:jjoosten@ub.edu}{jjoosten@ub.edu} }
\maketitle

\begin{abstract}
The principle of Generalized Natural Selection (GNS) states that in nature, computational processes of high computational sophistication are more likely to maintain/abide than processes of lower computational sophistication provided that sufficiently many resources are around to sustain the processes. In this paper we give a concrete set-up how to test GNS in a weak sense. In particular, we work in the setting of Cellular Automata and see how GNS can manifest itself in this setting. 
\end{abstract}

\section{Introduction}
One can perceive various physical processes in general and biological ones in particular as computational systems (see for example \cite{MembraneComputing}). By doing so, various notions and techniques of computer science become relevant and applicable in biological systems and vice versa.

In \cite{JoostenGNS} and \cite{JoostenFittest} Joosten  investigates these complexity notions to reflect on the direction of evolution in terms of complexity. A driving principle of Generalized Natural Selection (GNS) is proposed. GNS states that in nature, computational processes of high computational sophistication are more likely to maintain/abide than processes of lower computational sophistication provided that sufficiently many resources are around to sustain the processes.

A main ingredient of the argument is an analogy of Dawkins paradigm of the Selfish Gene (\cite{SelfishGene}) which in our setting could be called \emph{The Selfish Algorithm} as long as we do not interpret our analogy too strictly. In this paper we choose to model processes in nature by the parallel computing paradigm of Cellular Automata (CA) and study how GNS can be investigated in this setting. We present a methodology and some very preliminary results. As such the paper can be seen as a research roadmap to further investigations.

\section{A first implementation}

We want to design an experiment that captures, in a computer simulation, what happens when two different structures grow while both accessing a locally limited set of resources, and in particular when at least one of the structures presents some significant level of complexity. So, for this purpose we have chosen a one-dimensional cellular automaton (CA), with radius $1$ and three colors, white, grey and black which we denote by 0,1, and 2 respectively. 

Our CA works with a one-dimensional tape of discrete cells, where each cell can be of any of these three colors, and time evolution is modeled with discrete time steps. This way, we start by fixing an initial condition telling what cell is of what color. Our CA will use a rule that stipulates how to compute the color of a particular cell at a next time step depending on its current color and the color of its two direct neighbors. This way, the color of each cell will evolve over time.

More formally, we can label our tape-cells by $\ldots x_{-2}, x_{-1}, x_0, x_1, x_2 \ldots$. A tape-configuration is thus a map $c : \{ x_i \mid i \in \mathbb Z\} \to \{ 0,1,2\}$. We will denote the tape configuration at time $t$ ($t\in \mathbb N$) by $\ldots c_t(x_{-2}), c_t(x_{-1}), c_t(x_0), c_t(x_1) \ldots$. Thus, a CA with three colors in general is just a look-up table of  27 ($3\times 3\times 3$) rules telling you the value $c_{t+1}(x_n)$ depending on the triple\\ $\la c_t(x_{n-1}), c_t(x_{n}), c_t(x_{n+1})\ra$.

The main idea of this experiment is that black cells will correspond to a structure that follows some growing rule, grey cells correspond to a different structure that grows following another rule and white cells can be interpreted as a set of limited resources that black and grey cells use for their propagation. 

Thus, we shall restrict our attention to CAs with three colors which can be seen as composed by three CAs with two colors: one that determines the relation between white cells and black cells and fixes the growing of the \emph{black organism}, the analogous for the \emph{grey organism} and one that tells us what happens when cells of the different organisms contact in the same neighborhood. 

In particular we have that $c_{t+1}(x_n)$ should, be either white or black in case none of $c_{t}(x_{n-1})$, $c_{t}(x_{n})$ and $c_{t}(x_{n+1})$ is grey. Likewise, $c_{t+1}(x_n)$ should, be either white or grey in case none of $c_{t}(x_{n-1})$, $c_{t}(x_{n})$ and $c_{t}(x_{n+1})$ is black.

Admittedly one can argue that the correspondence between CAs and models of an organism leads way to discussion as to the ontological nature of what exactly \emph{is} the organism. In this paper we will not enter this discussion since the paper is a mere first exploration of what happens when complex behavior interacts with less complex behavior, or with behavior of a different type of complexity. So, deliberately we shall be rather vague as to the exact correspondence between black cells on the one hand and organisms or models of attributes thereof on the other hand.

Our CA model will be of such nature that from the contact between both colors only one of them survives or both disappear. This means that there are only three possible solutions to what happens when cells of both organisms contact in the same neighborhood: grey survives, black survives or there only remains a white cell. 

It is easy to check that there are exactly 12 possible neighborhoods (three adjacent cells) so that at least one black and at least one grey cell occur\footnote{From now on, we refer to this kind of neighborhoods as mixed neighborhoods.}.  In these cases , the values for the rule are assigned randomly so we do not interfere when both structures come in contact, giving priority to one over the other. This way, given any mixed neighborhood $N=\la c_t(x_{j-1}), c_t(x_j) , c_t(x_{j+1}) \ra, c_{t+1}(x_j)=r$ where $r$ is a random integer between 0 and 2. Since values are assigned randomly, we can expect that given $N=\la c_t(x_{j-1}), c_t(x_j) , c_t(x_{j+1}) \ra$ and $M=\la c_t(x_{i-1}), c_t(x_i) , c_t(x_{i+1}) \ra$, two different mixed neighborhoods, $c_{t+1}(x_j)$ and $c_{t+1}(x_i)$ might be different. For example, consider the neighborhood $\la 2,2,1 \ra$ and suppose that in this case, black cell survives, this does not mean that in the rest of the mixed neighborhoods, black cell survives, not even in similar cases as $\la 2, 1, 2 \ra$ or $\la 1, 2, 2 \ra$\footnote{This way, we are not considering totalistic rules that are the ones that depend only on the average color  of the cells in a neighborhood.}. Also, notice that there is one special neighborhood, $\la 0, 0, 0 \ra$ where all elements are white cells. For this case, we restrict the rule to return 0.

\section{Interacting CAs: some preliminary findings}
The authors are preparing a large simulation where interacting CAs are studied using the above methodology. Here, we report on some preliminary findings. As such we mention that we often see that the complexity of the resulting pattern indeed gets changed. Either we see that the more simple structure survives but increases its complexity or that the more complex structure survives while maintaining its complexity. 

Following Wolfram code, in the first two figures we can see respectively, Rules 90 and 110 running alone with just one black cell at the initial state. Rule 90 exhibits a nested pattern, while Rule 110 has a higher complexity. Moreover, Rule 110 it's known to be Turing complete.
\ \\

\begin{minipage}[b]{0.45\linewidth}
\centering \label{alone}
\includegraphics[width=\textwidth]{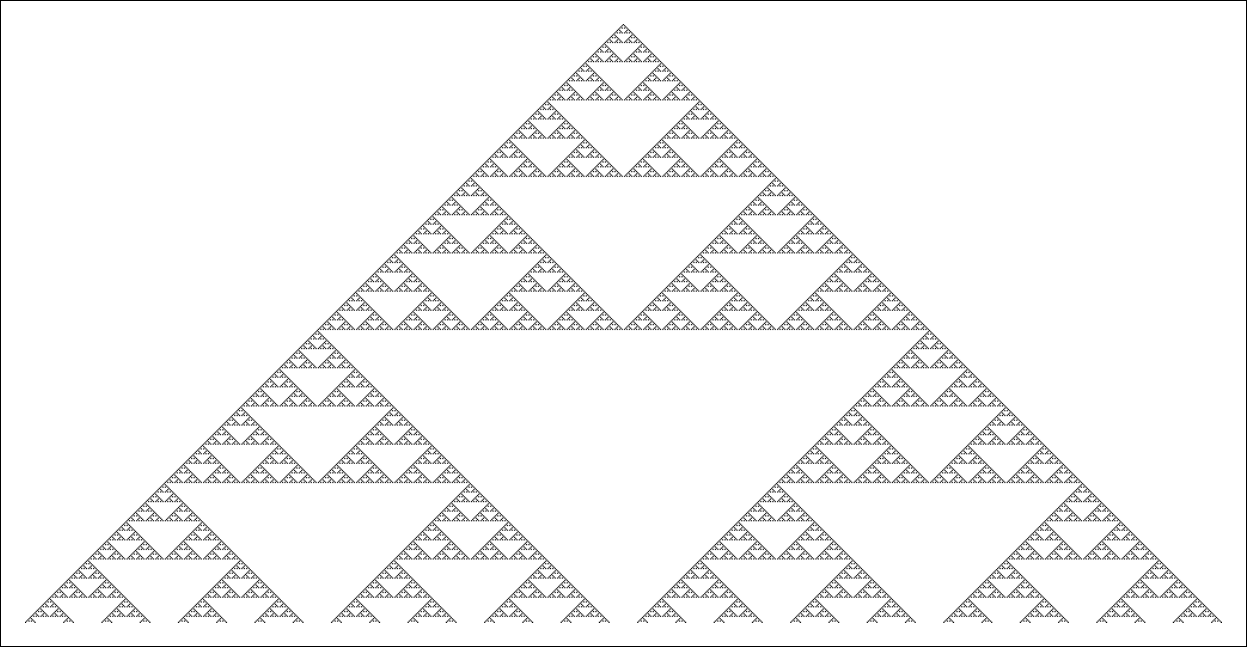}
\end{minipage}
\hspace{0.5cm}
\begin{minipage}[b]{0.45\linewidth}
\centering
\includegraphics[width=\textwidth]{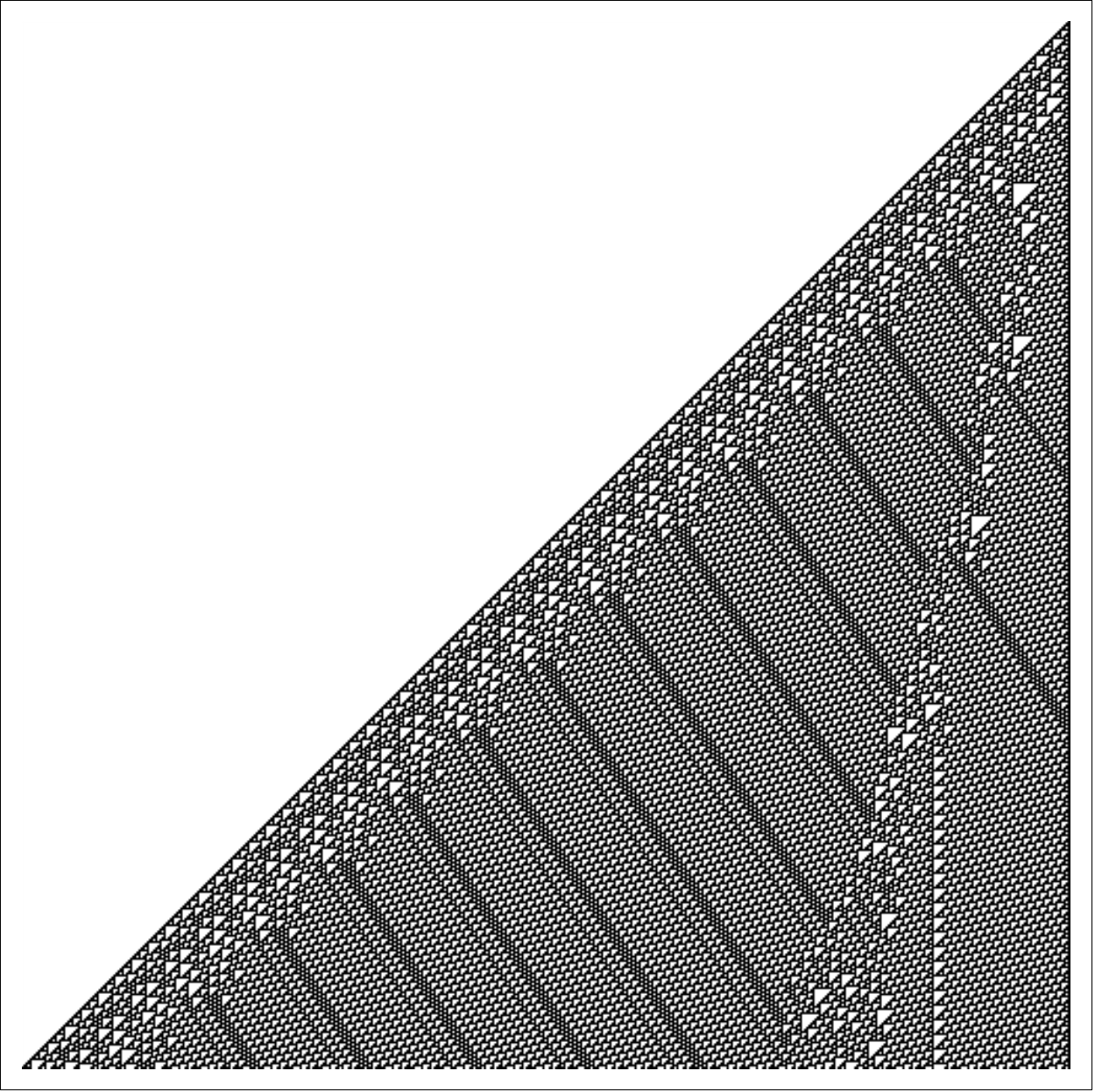}
\end{minipage}

\ \\

In the following figures we can see two different interaction between these cellular automata. In the first one, Rule 90 changes its behavior to a more complex one, generating a random substructure. Also, we can see that eventually Rule 110 disappears. In the second one, Rule 110 maintains its complexity and gains ground to Rule 90.

\ \\

\begin{minipage}[b]{0.45\linewidth}
\centering \label{special1}
\includegraphics[width=\textwidth]{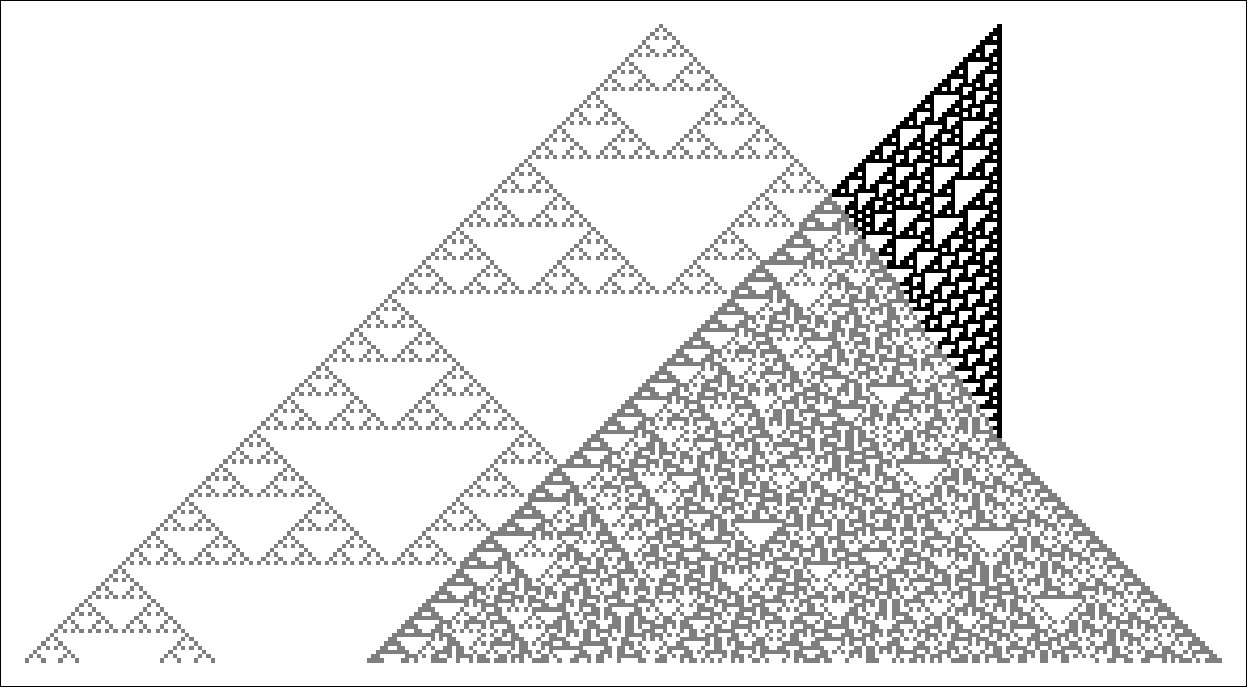}
\end{minipage}
\hspace{0.5cm}
\begin{minipage}[b]{0.45\linewidth}
\centering
\includegraphics[width=\textwidth]{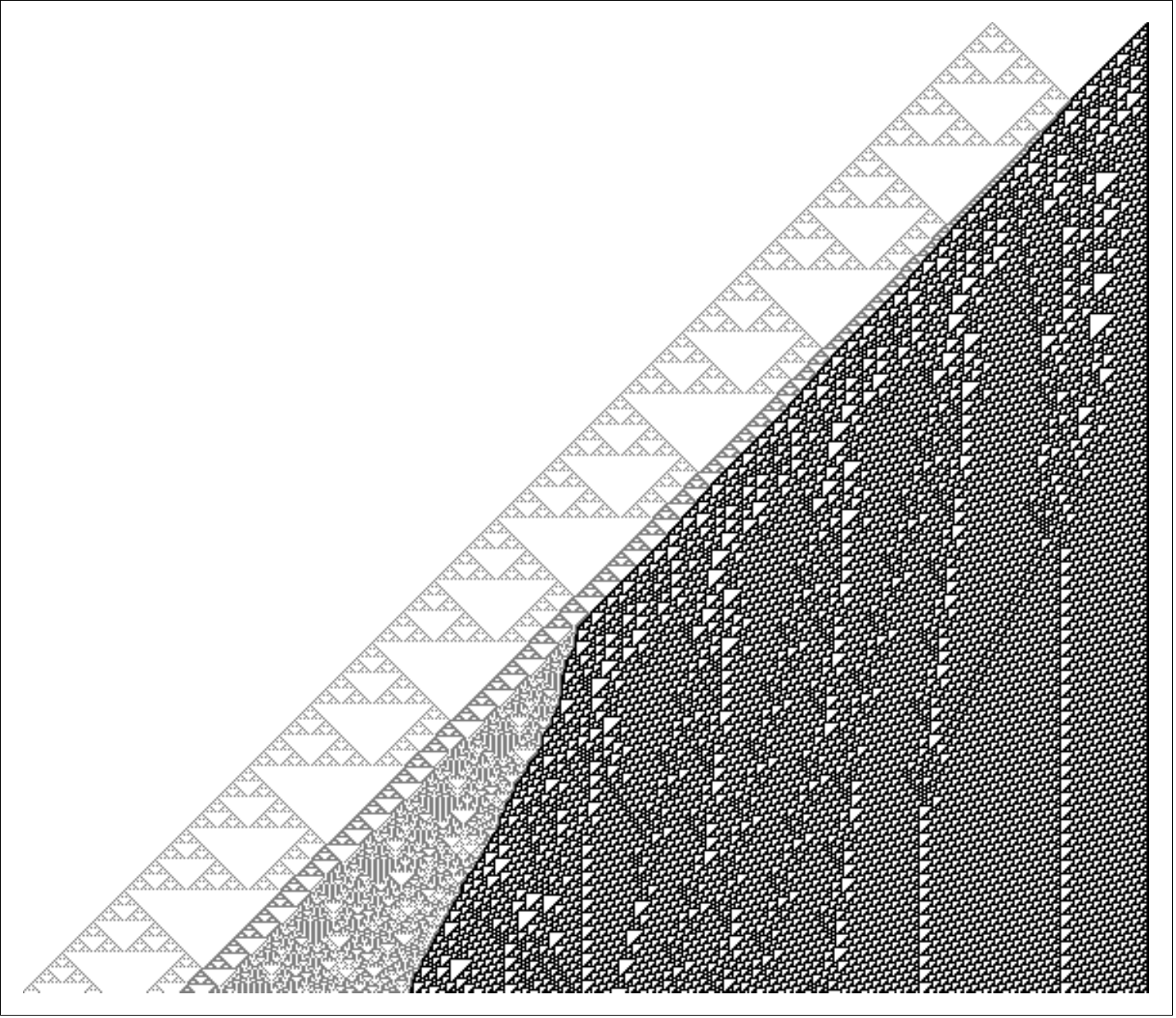}
\end{minipage}

\ \\

In forthcoming simulations an exhaustive experiment for interactions between all one-dimensional, two-colors, radius 1 CAs is planned. In this experiment, the complexity needs to be quantified, either by some entropy measure, or using some classification method. Moreover, the sensitivity for initial conditions needs to be taken into account by taking large samples.

\end{document}